\documentclass[conference]{IEEEtran}
\IEEEoverridecommandlockouts
\usepackage{cite}
\usepackage{amsmath,amssymb,amsfonts}
\usepackage{graphicx}
\usepackage{textcomp}
\usepackage{xcolor}
\usepackage{booktabs}
\usepackage{multirow}
\usepackage{url}  
\usepackage{algorithm}
\usepackage{algpseudocode}
\def\BibTeX{{\rm B\kern-.05em{\sc i\kern-.025em b}\kern-.08em
    T\kern-.1667em\lower.7ex\hbox{E}\kern-.125emX}}
\begin{document}

\title{AI-Enabled Digital Twins for Next-Generation Networks: Forecasting Traffic and Resource Management in 5G/6G
}

\author{\IEEEauthorblockN{1$^{st}$ John Sengendo}
\IEEEauthorblockA{Dept. of Information Engineering and Computer Science\\
University of Trento\\
Email: john.sengendo@unitn.it}
\and
\IEEEauthorblockN{2$^{nd}$ Fabrizio Granelli}
\IEEEauthorblockA{Dept. of Information Engineering and Computer Science\\
University of Trento\\
Email: fabrizio.granelli@unitn.it}
}

\maketitle

\begin{abstract}
As 5G and future 6G mobile networks become increasingly more sophisticated, the requirements for agility, scalability, resilience, and precision in real-time service provisioning cannot be met using traditional and heuristic-based resource management techniques, just like any advancing technology. With the aim of overcoming such limitations, network operators are foreseeing Digital Twins (DTs) as key enablers, which are designed as dynamic and virtual replicas of network infrastructure, allowing operators to model, analyze, and optimize various operations without any risk of affecting the live network.
However, for Digital Twin Networks (DTNs) to meet the challenges faced by operators especially in line with resource management, a driving engine is needed.
In this paper, an AI (Artificial Intelligence)-driven approach is presented by integrating a Long Short-Term Memory (LSTM) neural network into the DT framework, aimed at forecasting network traffic patterns and proactively managing resource allocation. Through analytical experiments, the AI-Enabled DT framework demonstrates superior performance benchmarked against baseline methods. Our study concludes that embedding AI capabilities within DTs paves the way for fully autonomous, adaptive, and high-performance network management in future mobile networks.

\end{abstract}


\section{Introduction}
The growing shift towards sophisticated mobile networks like 5G Advanced and next-generation networks has enabled new functionalities and ushered in unique challenges with respect to users' needs, demanding mobile heterogeneous services, and real-time resource allocation. Traditional management approaches are failing to achieve agility, intelligence, and responsiveness in the present day network environments as these studies, \cite{b1} \& \cite{b2} highlight. With excellence in areas such as the medical field where real-time monitoring and intelligent control are vital as noted in \cite{b3}, DTs are a new direction of focus in next-generation networks.

In the mobile networks ecosystem, Digital Twins are comprehensive, dynamic virtual replicas of actual networks operations that mirror topologies, components, and real-time operational states such as, traffic patterns.
By continuously ingesting live telemetry (traffic patterns, device metrics, etc.) and environmental inputs, DTs become “living” models that evolve in tandem with their physical counterpart \cite{b4}, \cite{b5}. This safe modeling environment provided by Digital Twins allows operators to analyze and optimize complex scenarios such as traffic surges, hardware failures, or service migrations without risking live-network disruptions \cite{b6}. Because DTs can execute hypothetical reconfiguration commands and incorporate contextual factors (e.g. device conditions or interference), they enable foresight into network behavior under change.
In practice, these twin-based capabilities support advanced planning and resource management (for example, forecasting capacity needs and re-allocating bandwidth to underutilized links) that traditional methods cannot achieve \cite{b7}.

Incorporating AI techniques into Digital Twins allows network operators to intelligently model and forecast resource behaviors as well as control their allocation in real-time, which improves the Quality of Service (QoS), energy consumption, and fault tolerance \cite{b2}, \cite{b8}. Additionally, AI-Enabled DTs allow operations to shift from a reactive response mode to proactive, and even fully autonomous operations which escalate operations like network segmentation, traffic forecasting, and resource provisioning, which require a high degree of precision and adaptability. This integration of AI forms the main topic of study in this paper, where we demonstrate how AI-driven approaches integrated within DTs excel in forecasting and network resource allocation.

The rest of the paper is organized as follows. In section II we present the related works in line with DTs and AI. In section III a detailed methodology of our work is presented. Section IV presents the results analysis from our study, and section V provides a conclusion of this work.

\section{Background and Related Works}

The rapid evolution of mobile networks is transforming how network infrastructures are deployed, managed, and optimized. These next-generation networks demand unprecedented agility, precision, and scalability to support latency-sensitive applications, including autonomous systems, remote healthcare, and immersive virtual reality. As studies in \cite{b9} suggest, static and traditional resource management strategies prevalent in legacy networks are increasingly inadequate to cope with the highly dynamic and heterogeneous traffic environment characteristic of 5G/6G ecosystems \cite{b9}.

Despite all of these challenges, Digital Twins (DTs) have  
emerged as a promising enabling technology. They offer  
virtualized and real-time representations, enabling continuous observance, and safe 
optimization of network operations without interrupting live  
services. As literature in \cite{b2} noted, DT-driven frameworks provide operational simulation and validation capabilities to resource  
management techniques before executing them in the real  
world, thus eliminating critical operational risks and network 
downtime \cite{b2}. It's thus, possible to consider the integration of DTs into network management pipelines as a  
self-sufficient cornerstone for future mobile networks.

Given the dynamic shifts in traffic patterns, enabling frameworks must incorporate highly accurate predictive engines focused on functional real-time resource allocation. AI implementations, particularly deep learning frameworks such as LSTM neural networks, as described in \cite{b10}, are suitable for this role due to their effectiveness in capturing complex traffic patterns and modeling temporal dependencies over significant periods of time \cite{b10}. Likewise, in studies presented by authors in\cite{b11}, they discussed strategies related to resource allocation to optimize IoT systems in 5G and 6G networks and stressed the degree of improvements offered by LSTM integration into the frameworks relative to traditional baseline approaches. Additionally, the study depicted in \cite{b12} shows that while generative AI coupled with DT networks can outperform non-generative LSTM models in high-load conditions, the latter strategy still retains its relevance in forecasting scenarios \cite{b12}. Their findings further strengthen the argument that hybrid AI-Driven DT solutions will dominate the discourse on 6G network management systems.

More studies such as in \cite{b13}, the authors investigated AI-Based DT frameworks for multimedia service provisioning, their findings show that the temporal modeling capacities of LSTM networks was highly effective for resource management tasks. Additional, more works supporting these findings is presented by authors in \cite{b14}, where they stress the scalability and accuracy of LSTM-directed configurations in challenging, multi-domain network configurations \cite{b14}.

Overall, current studies indicate an unmistakable research direction preferring AI-augmented DT models for intelligent network management. But as authors in \cite{b15} noted, there is, nevertheless, still quite a room to benchmark and show how AI-Driven DTs excel under realistic and variable traffic conditions. Filling this gap, in the subsequent sections we present the technical methodology of our work and later discuss results from our implementation framework. 

\section{Methodology}
In this section, we  provide a detailed methodology of our approach demonstrating the techniques performed and evaluation methods.

\subsection{Dataset Description}

\begin{figure*}[]
  \centering
  \includegraphics[width=\textwidth,height=12cm]{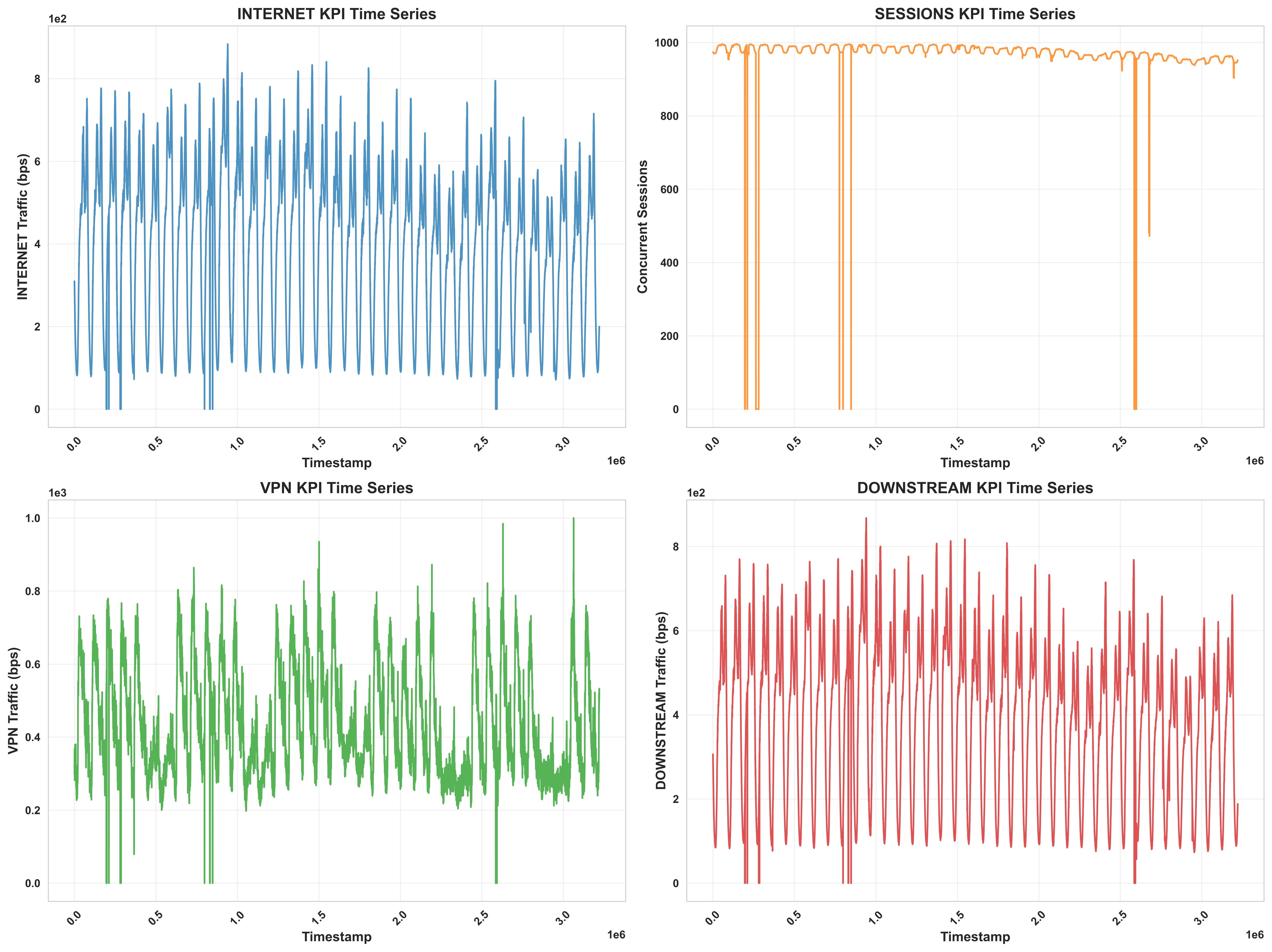}
  \caption{Dataset overview.}
  \label{fig:Data}
\end{figure*}
We used a time series dataset described in \cite{b16}, collected from a network operator. The dataset includes four key performance indicators (KPIs): \textit{internet usage}, \textit{active sessions}, \textit{VPN connections}, and \textit{downstream throughput}. The dataset includes time-series data tracking four essential network indicators sampled at 5-minute intervals over a period exceeding one month \cite{b16}. In Figure~\ref{fig:Data}, we provide a representation of the data showing the KPIs variations overtime. The data represents measurements from various operator facilities and encompasses: total Internet traffic flow (measured in bits per second), downlink traffic volume (in bits per second), count of concurrent user sessions, and VPN traffic throughput (in bits per second). To protect sensitive information, all data was anonymized, timestamps normalized to start from zero for each series, and metric values have been rescaled across all time series \cite{b16}.

In our LSTM approach following the standard sliding-window techniques, we used a fixed input window of past values to predict the future value of the \textit{internet} KPI. The final dataset consisted of an input array $X$ and a corresponding target vector $y$. $X$ represented as \( X \in \mathbb{R}^{n \times t \times 4} \), where:
\begin{itemize}
  \item \( n \) is the number of training samples (sliding windows).
  \item \( t \) is the number of time steps in the input window.
  \item The 4 features which are:
    \begin{enumerate}
      \item Total Internet traffic flow (bps)
      \item Downlink traffic volume (bps)
      \item Concurrent user sessions (count)
      \item VPN traffic throughput (bps)
    \end{enumerate}
\end{itemize}
Each element of the input tensor \( X_{i,j,k} \) represented: Sample \( i \), time step \( j \), feature \( k \in \{1,2,3,4\} \). The tensor is structured as below where each matrix slice in \( X \) represents a single training sample, capturing the temporal evolution of four features across \( t \) time steps.

\[
X = 
\left[
\begin{array}{c}
\left[
\begin{array}{cccc}
x^{(1)}_{1,1} & x^{(1)}_{1,2} & x^{(1)}_{1,3} & x^{(1)}_{1,4} \\
x^{(1)}_{2,1} & x^{(1)}_{2,2} & x^{(1)}_{2,3} & x^{(1)}_{2,4} \\
\vdots        & \vdots        & \vdots        & \vdots        \\
x^{(1)}_{t,1} & x^{(1)}_{t,2} & x^{(1)}_{t,3} & x^{(1)}_{t,4} \\
\end{array}
\right] \\
\\
\left[
\begin{array}{cccc}
x^{(2)}_{1,1} & x^{(2)}_{1,2} & x^{(2)}_{1,3} & x^{(2)}_{1,4} \\
x^{(2)}_{2,1} & x^{(2)}_{2,2} & x^{(2)}_{2,3} & x^{(2)}_{2,4} \\
\vdots        & \vdots        & \vdots        & \vdots        \\
x^{(2)}_{t,1} & x^{(2)}_{t,2} & x^{(2)}_{t,3} & x^{(2)}_{t,4} \\
\end{array}
\right] \\
\vdots \\
\left[
\begin{array}{cccc}
x^{(n)}_{1,1} & x^{(n)}_{1,2} & x^{(n)}_{1,3} & x^{(n)}_{1,4} \\
x^{(n)}_{2,1} & x^{(n)}_{2,2} & x^{(n)}_{2,3} & x^{(n)}_{2,4} \\
\vdots        & \vdots        & \vdots        & \vdots        \\
x^{(n)}_{t,1} & x^{(n)}_{t,2} & x^{(n)}_{t,3} & x^{(n)}_{t,4} \\
\end{array}
\right]
\end{array}
\right]
\]

The corresponding target vector \( y \in \mathbb{R}^{n} \) contained the future value of the Internet KPI (e.g., total traffic flow) represented as:

\[
y =
\begin{bmatrix}
y^{(1)} \\
y^{(2)} \\
\vdots \\
y^{(n)}
\end{bmatrix}
\]
The dataset was split chronologically into training, validation and test set without shuffling to preserve temporal integrity in time series.

\subsection{Baseline Methods}
\begin{algorithm}[H]
\caption{Baseline 1 ($\mu + 2\sigma$) Forecasting }
\begin{algorithmic}[1]
\Require Training series $Y_{\text{train}} = \{y_1, y_2, \ldots, y_T\}$, Forecast horizon $N$
\Ensure Forecast $\hat{Y} = \{\hat{y}_1, \hat{y}_2, \ldots, \hat{y}_N\}$

\State Compute $\mu \gets \frac{1}{T} \sum_{t=1}^{T} y_t$
\State Compute $\sigma \gets \sqrt{\frac{1}{T} \sum_{t=1}^{T} (y_t - \mu)^2}$
\State $c \gets \mu + 2\sigma$
\For{$i = 1$ to $N$}
    \State $\hat{y}_i \gets c$
\EndFor
\State \Return $\hat{Y}$
\end{algorithmic}
\end{algorithm}
In the framework of \textbf{Baseline 1}, we adopted a static
procedure based on the mean and standard deviation
of \texttt{internet} KPI values from the training dataset.
As illustrated in Algorithm 1, we calculated the
mean and then added twice the standard deviation to account
for variability and probable peak demands. This computation
was then used to act as a baseline prediction during subsequent future time steps during texting. This baseline approach provided an upper-bound estimate, to refer to as a reference
point for comparison with other more dynamic prediction models.
\begin{algorithm}[H]
\caption{Baseline 2 Percentile (P95) Forecasting}
\begin{algorithmic}[1]
\Require Training series $Y_{\text{train}} = \{y_1, y_2, \ldots, y_T\}$, Forecast horizon $N$
\Ensure Forecast $\hat{Y} = \{\hat{y}_1, \hat{y}_2, \ldots, \hat{y}_N\}$

\State Compute $p_{95} \gets \text{Percentile}(Y_{\text{train}},\ 95)$
\For{$i = 1$ to $N$}
    \State $\hat{y}_i \gets p_{95}$
\EndFor
\State \Return $\hat{Y}$
\end{algorithmic}
\end{algorithm}
For a more comparative approach, we also implemented \textbf{Baseline 2} which adopted a percentile-based strategy. In this method, as we illustrate in Algorithm 2 we computed the 95th percentile (P95) of the \texttt{internet} KPI values from the training set and used this outcome to forecast all future time points. This approach was adopted to capture high-usage scenarios by focusing on extreme values observed in the historical data. The P95 baseline is particularly useful in bandwidth provisioning, as it ensures that allocated resources exceed the demand in 95\% of the training instances \cite{b17}, offering a robust alternative to the mean-based baseline approach.

\subsection{AI Enabled Digital Twin Model: Bidirectional LSTM}

We modeled our AI-DT approach on a deep bidirectional long-short-term memory (BiLSTM) network, which is capable of capturing complex temporal dependencies in multivariate sequences \cite{b18,b19} where \( \mathbf{X} = \{ \mathbf{x}_1, \mathbf{x}_2, \ldots, \mathbf{x}_T \} \) denoted the input sequence, where each \( \mathbf{x}_t \in \mathbb{R}^d \) represented a \(d\)-dimensional feature vector at time step \(t\). The BiLSTM processes the input in both forward and backward directions as elaborated in \cite{b19}, producing hidden states \( \overrightarrow{\mathbf{h}}_t \) and \( \overleftarrow{\mathbf{h}}_t \), which are then concatenated to form the context representation \( \mathbf{h}_t = [\overrightarrow{\mathbf{h}}_t; \overleftarrow{\mathbf{h}}_t] \).

The specific layer configurations, regularization strategy (including dropout rate \( p \)), and optimizer settings (such as Adam with learning rate \( \eta \)) where set as depicted in Table~\ref{tab:exp-setup} above. The final output of the model being a scalar \( \hat{y}_{T+1} \in \mathbb{R} \), representing the predicted internet demand at the next time step \( T+1 \), computed as:
\[
\hat{y}_{T+1} = f_{\theta}(\mathbf{X}),
\]
where \( f_{\theta} \) denotes the BiLSTM model parameterized by \( \theta \).

Training was conducted using a mean squared error (MSE) loss function,
\[
\mathcal{L}(\theta) = \frac{1}{N} \sum_{i=1}^{N} \left( y_{T+1}^{(i)} - \hat{y}_{T+1}^{(i)} \right)^2,
\]
where \( N \) is the number of training samples. Algorithm 3 further demonstrates the work flow of this approach. 
\begin{table}[t]
\centering
\scriptsize
\caption{Summary of Experimental Setup}
\label{tab:exp-setup}
\begin{tabular}{|p{3.2cm}|p{4.9cm}|}
\hline
\textbf{Category} & \textbf{Details} \\
\hline
\textbf{Data \& Preprocessing} &
KPIs: Internet usage, sessions, VPNs, downstream throughput (Mbps). \\
& Target: Internet usage at \(t+1\) (Mbps); window: 10-step \(\rightarrow\) 1-step-ahead. \\
& Tensor: \([N,10,4]\); Split: 80\% train / 20\% test; 20\% validation from training. \\
\hline
\textbf{Model Architecture} &
BiLSTM (128) \(\rightarrow\) BiLSTM (128) \(\rightarrow\) BiLSTM (64) \(\rightarrow\) Dense (1). \\
& Dropout (\( p \) = 0.2), BatchNorm; Optimizer: Adam + LR scheduler. \\
\hline
\textbf{Training Settings} &
Max 30 epochs; early stopping (patience = 5); Batch size: 32; LR: 0.0001. \\
\hline
\textbf{Baselines} &
\textbf{B1:} \(\mu_{y_{\text{train}}} + 2\sigma_{y_{\text{train}}}\). \\
& \textbf{B2:} 95th percentile \(P_{95}(y_{\text{train}})\). \\
\hline
\textbf{Metrics} &
MAE, RMSE, Efficiency, Wastage, Utilization, Over-Provisioning. \\
\hline
\end{tabular}
\end{table}

\subsection{Evaluation Metrics}

We evaluated both the accuracy and operational efficiency of bandwidth allocation using the following metrics.
For effectiveness of the proposed approaches, two main categories of metrics are employed:

\begin{itemize}
    \item \textbf{Forecasting Accuracy}: The predictive performance was quantified using the Mean Absolute Error (MAE) and the Root Mean Squared Error (RMSE), defined as:
    \begin{equation}
        \text{MAE} = \frac{1}{n} \sum_{i=1}^{n} \left| \hat{y}_i - y_i \right|, \quad
        \text{RMSE} = \sqrt{ \frac{1}{n} \sum_{i=1}^{n} \left( \hat{y}_i - y_i \right)^2 }
    \end{equation}
    where \( \hat{y}_i \) and \( y_i \) denote the predicted and true values at time step \( i \), respectively.

    \item \textbf{Allocation Performance}: The effectiveness of the resource allocation strategy was evaluated using the following metrics:
    \begin{equation}
        \text{Efficiency} = \min\left(\frac{\text{actual}}{\text{allocated}}, 1\right)
    \end{equation}
    \begin{equation}
        \text{Wastage} = \max\left(\frac{\text{allocated} - \text{actual}}{\text{allocated}}, 0\right)
    \end{equation}
    \begin{equation}
        \text{Utilization} = \frac{\text{actual}}{\text{allocated}}
    \end{equation}
    \begin{equation}
        \text{Over\text{-}Provisioning} = \max(\text{allocated} - \text{actual}, 0)
    \end{equation}
\end{itemize}

where \textit{actual} represented the true resource demand, and \textit{allocated} denoted the predicted resource allocation. These metrics captured different aspects of allocation effectiveness, including under-utilization, efficiency, and potential over-provisioning.

\begin{algorithm}[]
\caption{Bidirectional LSTM for Time Series Forecasting}
\begin{algorithmic}[1]
\Require Multivariate series $X \in \mathbb{R}^{T \times d}$, window size $w$, target index $j$
\Ensure Forecasts $\hat{y}_{w+1}, \ldots, \hat{y}_{T}$

\Statex \textbf{Preprocessing}
\For{$t = 1$ to $T - w$}
    \State $X^{(t)} \gets X_{t:t+w-1}$ \Comment{Input window}
    \State $y^{(t)} \gets X_{t+w,\ j}$ \Comment{Target value}
\EndFor
\State Split into train, validation and test sets: $\{X^{(t)}, y^{(t)}\} \rightarrow \mathcal{D}_{\text{train}}, \mathcal{D}_{\text{test}}$

\Statex \textbf{Model}
\State Define Bidirectional LSTM with 3 layers: [128, 128, 64 units], with dropout and batch norm
\State Final layer: Dense(64) $\rightarrow$ Dense(1)

\Statex \textbf{Training}
\State Train model on $\mathcal{D}_{\text{train}}$ with Adam optimizer and MSE loss

\Statex \textbf{Inference}
\For{each $X^{(t)} \in \mathcal{D}_{\text{test}}$}
    \State $\hat{y}^{(t)} \gets \text{model}(X^{(t)})$
\EndFor
\State \Return $\{\hat{y}^{(t)}\}$
\end{algorithmic}
\end{algorithm}

\section{Results Discussion}
\begin{figure}[htbp]
  \centering
  \includegraphics[width=\columnwidth, height=5cm]{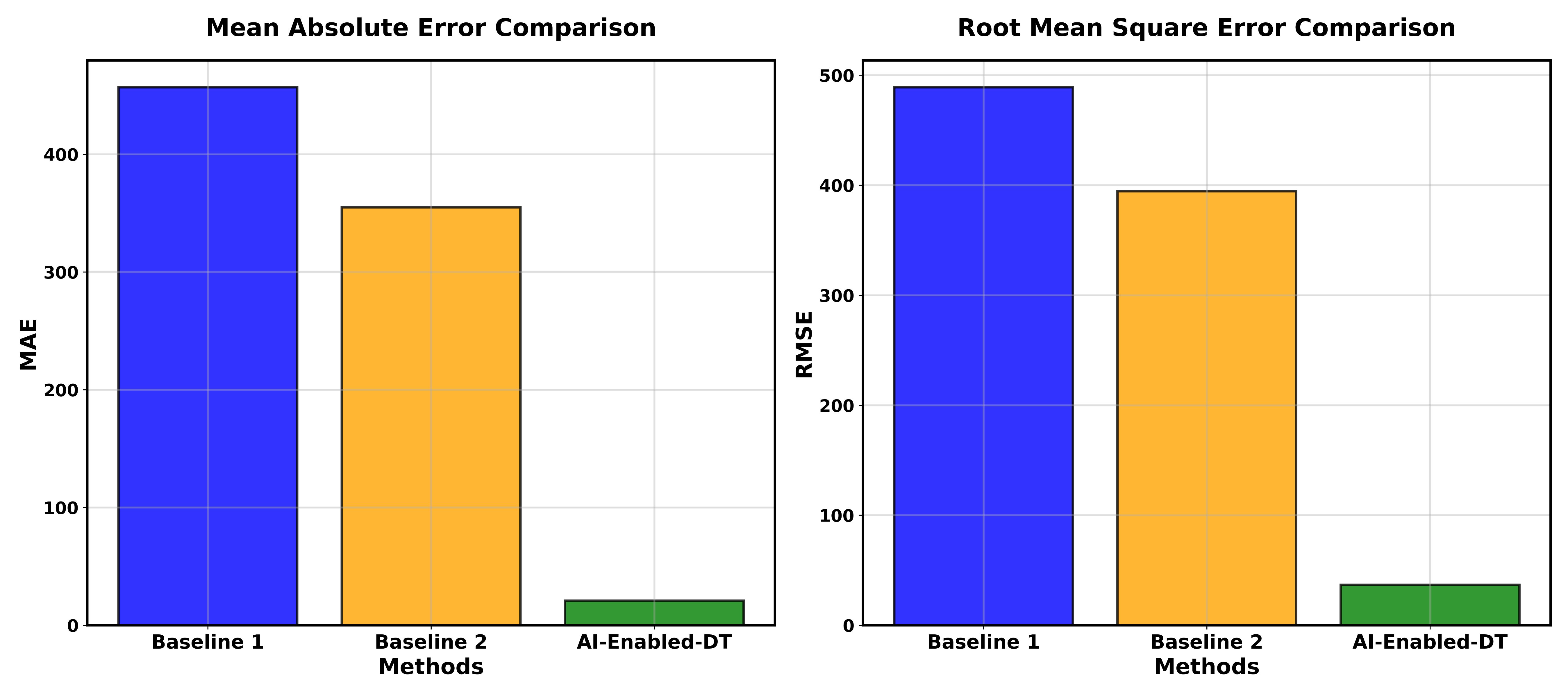}
  \caption{MAE and RMSE comparison.}
  \label{fig:MAE}
\end{figure}
\begin{figure}[htbp]
  \centering
  \includegraphics[width=\columnwidth, height=6cm]{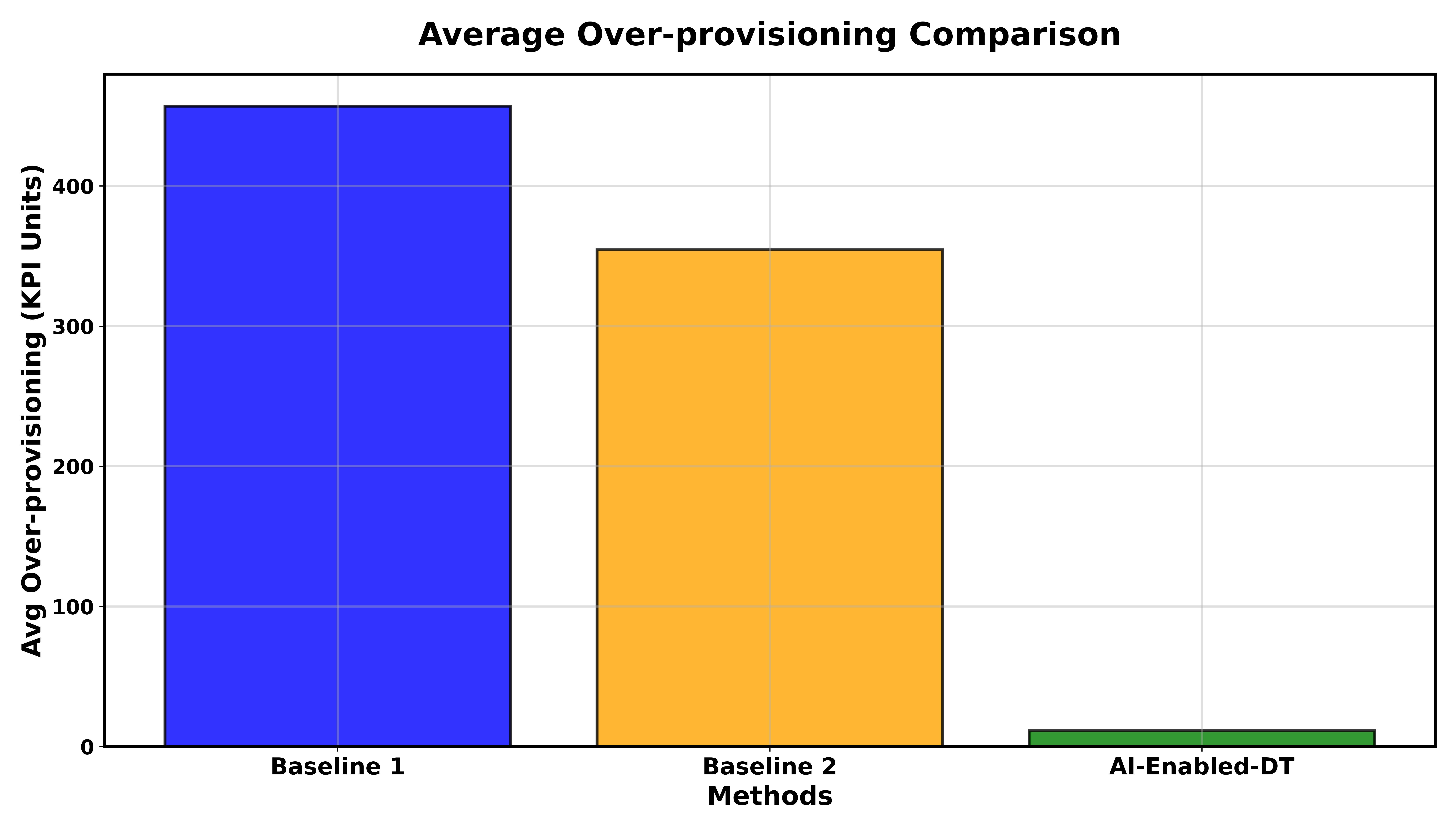}
  \caption{Over provisioning comparison.}
  \label{fig:OVER}
\end{figure}

In this section we provide an analysis and visual representation of the results obtained. Figures~\ref{fig:MAE} and~\ref{fig:OVER} provide a quantitative comparison between
the proposed AI-Enabled Digital Twin strategy to the
two baseline approaches among three prominent performance metrics: Mean Absolute Error, Root Mean Square Error
, and the mean over-provisioning respectively. In forecasting accuracy, the AI-Enabled DT approach achieves much lower error values as the green bar depicts, with
approximately $\text{MAE} \approx 25$ and $\text{RMSE} \approx 40$, while Baseline 1 and Baseline 2 demonstrated very large errors.

On average over-provisioning analysis (in terms of KPI
units), the AI-Enabled DT exhibited excellent performance with a significant lower value, benchmarked against Baseline 1 and Baseline 2 which
show values higher than 450 and 350 units, as shown in Figures~\ref{fig:OVER} below. This
highlighted the capacity of the DT not only to enhance predictive
accuracy through a lower prediction/forecasting error but also eliminating resource allocation inefficiencies in
dynamic network environments.

\begin{figure}[htbp]
  \centering
  \includegraphics[width=\columnwidth, height=6cm]{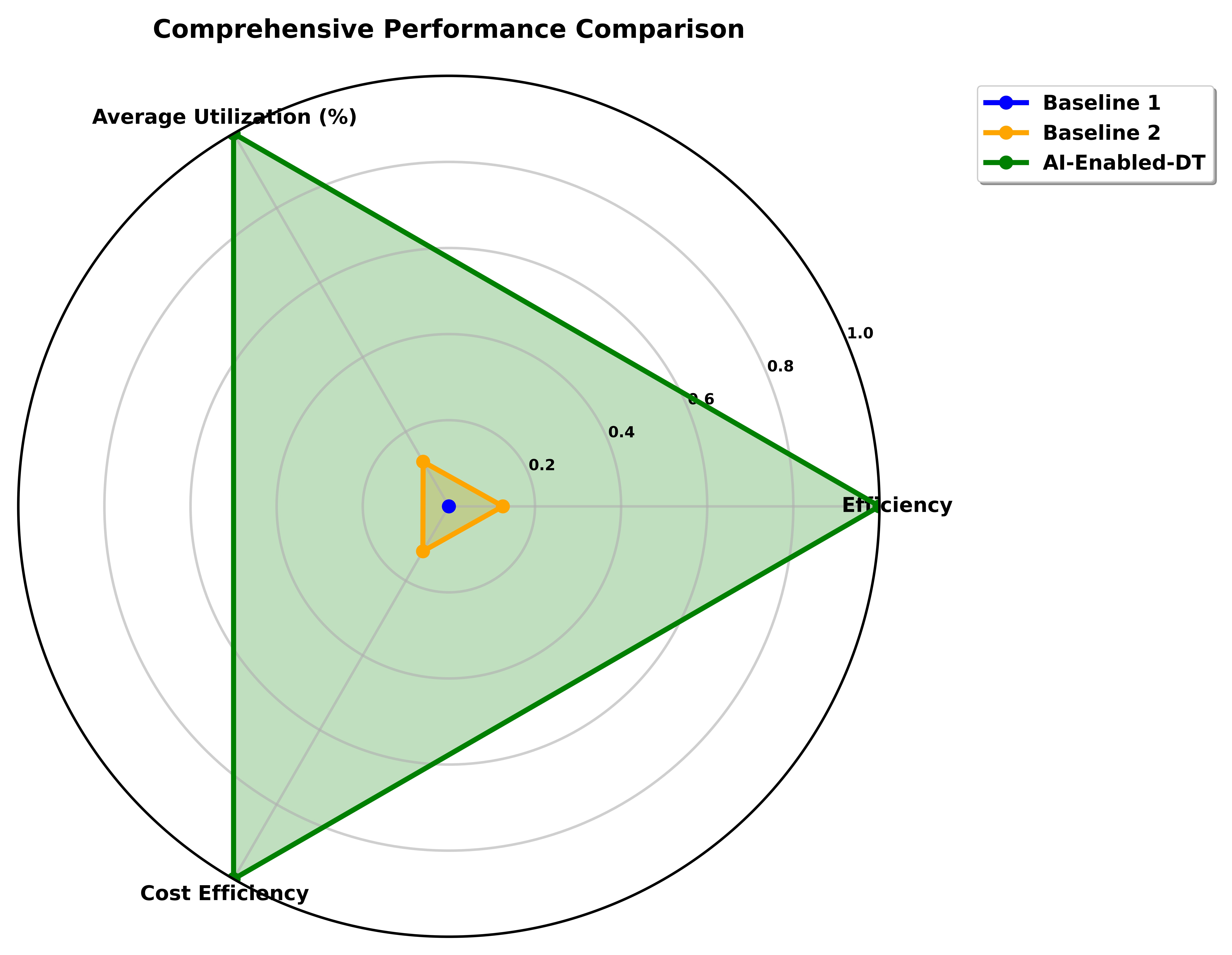}
  \caption{Radar chart.}
  \label{fig:rada}
\end{figure}
\begin{figure}[htbp]
  \centering
  \includegraphics[width=\columnwidth, height=5cm]{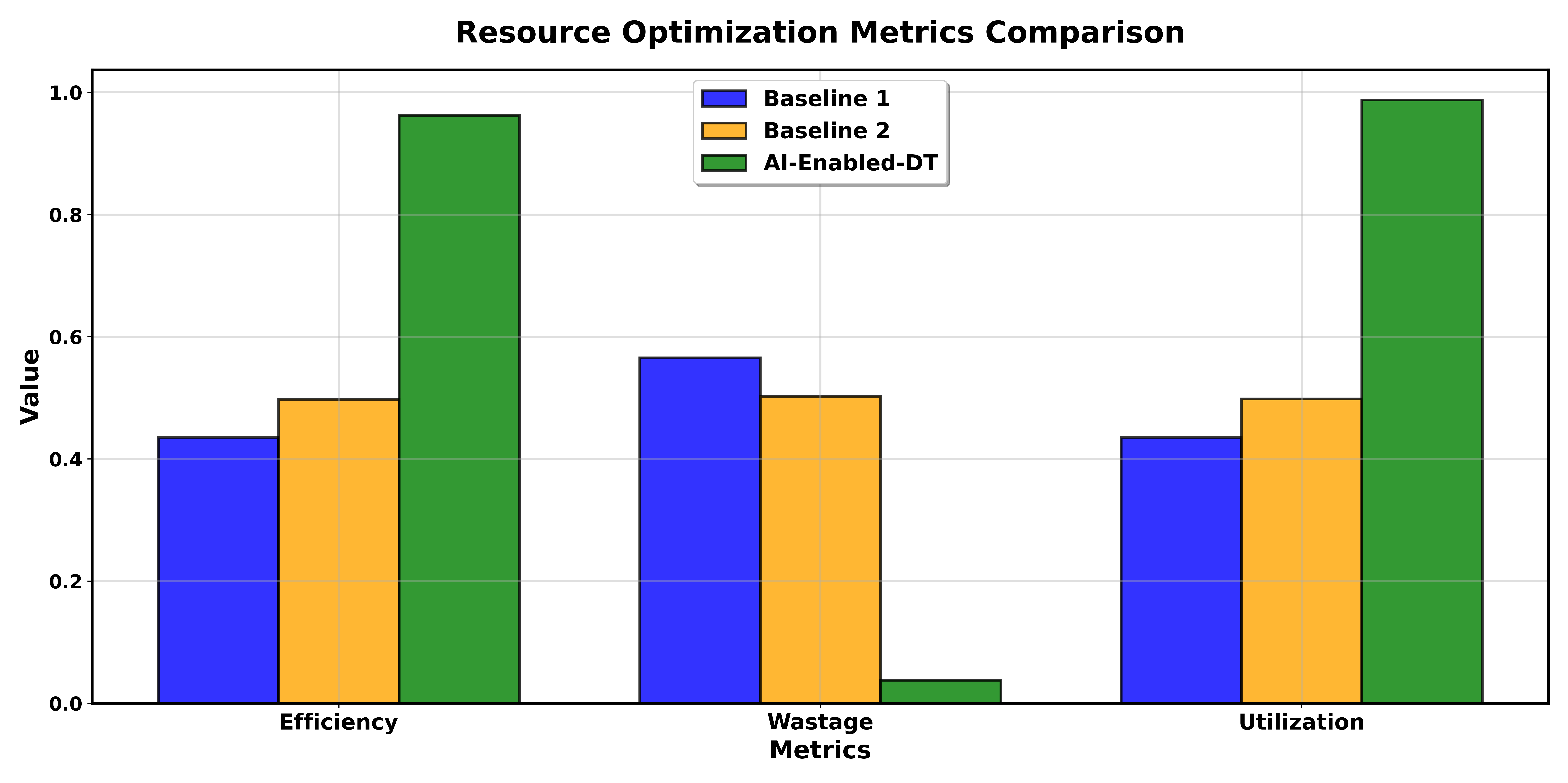}
  \caption{Resource optimization comparison.}
  \label{fig:adva}
\end{figure}

The radar chart we present in Figure 4 provides a visualization
of performance where all metric values are normalized on a
scale from 0 to 1, where higher values indicate better performance.
As depicted, AI-Enabled DT method achieved near-maximum
normalized performance across all evaluated dimensions, with
particularly strong performance in efficiency ($\eta_{\text{norm}} \approx 1.0$), cost efficiency ($CE_{\text{norm}} \approx 1.0$) and average utilization. These results indicate
that the proposed AI-Enabled approach not only provides
superior predictive accuracy but also enables significant resource
optimization, reducing operational costs.

In terms of resource optimization metrics as depicted in Figure~\ref{fig:adva}, the AI-Enabled DT method achieves an efficiency $\eta \approx 0.95$, indicating near-optimal resource utilization, while maintaining minimal wastage ratios ($\omega < 0.05$). This advanced metrics comparison illustrates that the AI-Enabled DT approach exhibits an efficient utilization factor ($\rho \approx 1$), significantly outperforming both baselines when it comes to efficient optimization of network resources.

\begin{figure}[htbp]
  \centering
  \includegraphics[width=\columnwidth, height=6cm]{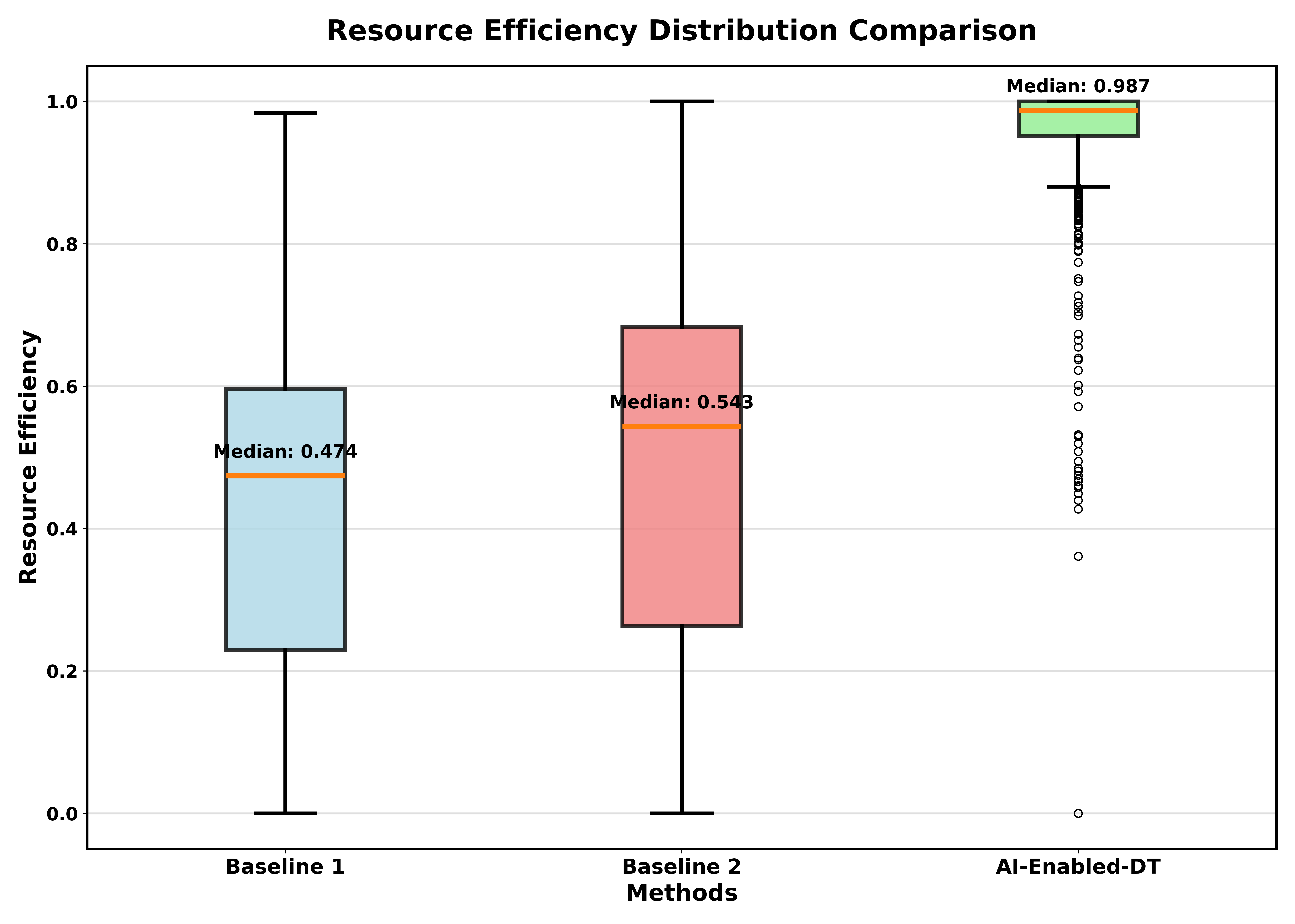}
  \caption{Resource distribution comparison.}
  \label{fig:Re}
\end{figure}

Additionally, in Figure~\ref{fig:Re} we present a comparative analysis of resource efficiency distributions across the two baselines and the proposed AI-Enabled DT model. As observed, Baseline 1 had a median resource efficiency of 0.474, while Baseline 2 showed a slightly higher median of 0.543. In contrast, the AI-enabled DT significantly outperforms both baselines with a median efficiency of 0.987. Additionally, the AI-Enabled DT exhibited lower variability and a more compact inter-quartile range, indicating more consistent performance. These results highlight the effectiveness of integrating AI-based forecasting and control into the DT framework for optimizing resource usage.

\begin{table}[h]
\centering
\caption{Model size and In-memory comparison across methods}
\resizebox{\columnwidth}{!}{%
\begin{tabular}{|l|c|c|c|}
\hline
\textbf{Metric} & \textbf{Baseline 1} & \textbf{Baseline 2} & \textbf{AI-Enabled DT} \\
\hline
Model Size (MB) & 2.00 & 1.00 & 2.69 \\
\hline
In-memory Requirement & Moderate & Lightweight & Heavy \\
\hline
\end{tabular}%
}
\label{tab:model_size}
\end{table}

In line with computational complexity, we provide results in Tabel II which indicate memory footprint requirement by each method. Our results indicated that the Baseline models are more light weight in comparison to the AI-Driven DT. Despite the excellent performance in forecasting, resource management and optimization provided by the AI-Enabled DT framework, there still a challenge of computational resource requirement both for memory requirement and computation complexity as there are many training parameters needed as it was shown in Tabel I. This creates an open research direction of solving the price needed to pay for computation as also highlighted in recent studies like \cite{b20}.

\section{conclusion}
 In this paper, we presented a comparative study that demonstrated both traditional techniques and an AI-Enabled Digital Twins framework in optimizing resource management. Our findings showed how superior AI enabling can improve forecasting and resource management, highlighting the potential for improved network management. We propose that enabler technologies like AI-Enabled Digital Twins are fundamental in sustaining reliability and efficiency of next-generation mobile networks like 6G where real-time monitoring and optimization is fundamental. Future research work will focus on computational requirement of AI, so that network service providers can benefit from both efficient resource management and less computational requirement.
\section*{Acknowledgment}
This work has been carried out with support from the HORSE project (Holistic, Omnipresent, Resilient Services for Future 6G Wireless and Computing Ecosystems), funded by the Smart Networks and Services Joint Undertaking (SNS JU).

\end{document}